\documentclass[12pt]{article}
\usepackage{amssymb}
\usepackage{times}
\usepackage[nointegrals]{wasysym}
\usepackage{epsf,amsmath, amsfonts}
\usepackage{multirow,multicol,chemarr}
\usepackage{graphicx,subfigure}
\usepackage{dcolumn}
\usepackage{bm}
\usepackage{epsfig,latexsym,wrapfig}
\usepackage{pstricks}
\usepackage[activate={true,nocompatibility},final,tracking=true,kerning=true,spacing=true,factor=1100,stretch=10,shrink=10]{microtype}
\usepackage{chemarr,color,nicefrac}
\usepackage[top=1in, bottom=1in, left=1in, right=1in]{geometry}

\newcommand{\ben}{\begin{equation}}
\newcommand{\een}{\end{equation}}
\newcommand{\bea}{\begin{array}}
\newcommand{\eea}{\end{array}}
\newcommand{\bef}{\begin{figure}}
\newcommand{\eef}{\end{figure}}

\date{ }

\begin{document}
\title{\large \bf Ion mediated crosslink driven mucous swelling kinetics} 
\author{ {\bf S. Sircar$^a$ and A. J. Roberts}\\{\small School of Mathematical Sciences, University of Adelaide, SA 5005, Australia}\\{\small $^a$ Corresponding Author (email: sarthok.sircar@adelaide.edu.au)} }
\maketitle

\begin{abstract}
We present an experimentally guided, multi-phasic, multi-species ionic gel model to compare and make qualitative predictions on the rheology of mucus of healthy individuals (Wild Type) versus those infected with Cystic Fibrosis. The mixture theory consists of the mucus (polymer phase) and water (solvent phase) as well as several different ions: H$^+$, Na$^+$ and Ca$^{2+}$. The model is linearized to study the hydration of spherically symmetric mucus gels and calibrated against the experimental data of mucus diffusivities. Near equilibrium, the linearized form of the equation describing the radial size of the gel, reduces to the well-known expression used in the kinetic theory of swelling hydrogels. Numerical studies reveal that the Donnan potential is the dominating mechanism driving the mucus swelling/deswelling transition. However, the altered swelling kinetics of the Cystic Fibrosis infected mucus is not merely governed by the hydroelectric composition of the swelling media, but also due to the altered movement of electrolytes as well as due to the defective properties of the mucin polymer network.
\end{abstract}


\section{Introduction} \label{sec:intro}
\begin{wrapfigure}{r}{0.4\textwidth}
\vskip -5pt
  \centering
  \includegraphics[width=0.4\textwidth]{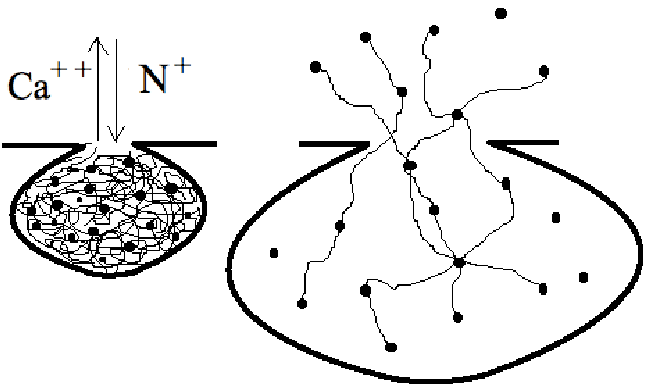}
 \caption{Schematics showing how the ion-displacement changes between calcium ion (Ca$^{2+}$) and a monovalent ion (e.g., Na$^+$) leading to changes in the network structure of the mucus matrix and its eventual expansion.}\label{fig:Fig1}
\end{wrapfigure}

Mucus is a polyelectrolyte biogel that plays a critical role as a protective, exchange and transport medium in the digestive, respiratory and reproductive systems of humans and other vertebrates~\cite{Verdugo1984,Verdugo1993}. Its swelling mechanism are of special interest because of its role in understanding a variety of diseases including cystic fibrosis (CF)~\cite{Barasch1991,Cheng1989,Kuver2000}. The conformation of the long-chain, negatively charged mucus glycoproteins depends strongly on factors such as pH, ionic strength and ionic bath composition \cite{Bansil2006}. Mucin is present in secretory vesicles at very high concentrations where they  are shielded primarily by a combination of divalent ions (e.g., Ca$^{2+}$) \cite{Villar2007}. Experiments show  that the mucus gel may swell explosively, up to 600-fold, in times that are the order of a few seconds, a process not observed in hydrogel swelling \cite{Verdugo1998, Verdugo1987}. Experiments also confirm that this rapid and massive expansion of the mucus gel is driven by an exchange of calcium in the vesicle for a monovalent ion (e.g., Na$^+$) in the extracellular environment \cite{Verdugo1987}. This is because, calcium being divalent, must balance two negative charges rather than one. Hence, a divalent Ca$^{2+}$ ion can act as a `cross-linker' between two polymer strands, allowing much tighter condensation than when the negative charges of the network are shielded by monovalent ions (Fig.~\ref{fig:Fig1}). Further, experiments demonstrate that the exocytosed mucin is recondensed if the calcium concentration of  the ionic medium is increased sufficiently \cite{Baconnais2005}. These observations indicate that the amount and the nature of the salt dissolved in the solvent determines the initial and the equilibrium configuration of these gels.

Experimental findings have reported that (unlike hydrogel swelling) a simple osmotic pressure difference of solute particles across the mucus gel cannot explain the massive and the explosive post-exocytotic swelling observations \cite{Tam1981}. Mucus hydration results from a balance between osmotic forces, diffusional hindrance (of the {\em ensemble of entangled polymer mesh} of the mucus), polyionic charges of the mucin and the other fixed polyions entrapped within the gel's matrix, the last two factors being the main idea behind the forces via Donnan potential \cite{Katchalsky1951}. The Donnan potential is not constant, but modulated by the concentration of free cations and polycations and the pH of the hydrating fluid \cite{Wolgemuth2004}. Marriott et al. observed that a decreased monovalent ion concentration, or an increased Ca$^{2+}$ concentration, in the airway surface liquid was a limiting factor in mucus hydration in CF-infected mucus \cite{Crowther1983}. Lisle indicated the role of defective processing of mucin polymers, via increased sulfation, in the abnormal mucus hydration in CF \cite{Lisle1995}. Increased sulfation and sialilation found in CF mucins \cite{Chase1983} are important for mucus swelling, since the high affinity of sulfate residues for calcium drastically shifts the normal monovalent/divalent ion-exchange properties of mucins and their swelling characteristics following the exocytotic release \cite{Barash1991}. In summary, besides the ionic composition of water on the surface of the airway, the defective processing of mucins and its abnormal ion exchange properties are essential factors to explain the characteristically deficient hydration and rheology of mucus in CF. However, the precise quantitative link between these elements and mucus hydration still remains unknown~\cite{Verdugo_NATURE1998}.

The theory to understand the  swelling and deswelling of ionic gels has a long history beginning 
with the classical work of Flory \cite{Flory1953,Flory1943a,Flory1943b} and Katchalsky \cite{Katchalsky1955} (see also \cite{Doi1986,Flory1976}). An early study to understand the kinetics (and not simply the equilibria) of swelling and deswelling, was done by Tanaka and colleagues \cite{Tanaka1979,Tanaka1980} who developed a kinetic theory of swelling gels viewing a  gel as a linear elastic solid immersed in a viscous fluid.  Although they neglected the motion of the fluid solvent, the model reasonably explained the swelling of a gel to its equilibrium volume fraction. Their work gave rise to the concept of gel diffusivity as a way to characterize the kinetics of swelling. The gel diffusivity is defined as $D = \nicefrac{L^2}{\tau}$ where $L$ is the equilibrium size (length or radius) of a gel and $\tau$ is the time constant of exponential swelling toward the equilibrium size. They found that expansion of the gel was governed approximately by a diffusion equation with diffusion coefficient $D$. However, how the diffusivity of the gel, or more generally the kinetics of swelling (perhaps with large changes in volume fraction) is affected by the movement of the ions and defective properties of the mucus polymer network is poorly understood.

 Subsequent studies relaxed the assumption of linear elasticity by defining the force on the gel 
 to be the functional derivative of the free energy function for the polymer mesh \cite{Durning1993, Maskawa1999, Onuki1989, Sekimoto1989, Yamaue2000}. However, most of these works neglected the fluid flow that must accompany swelling. Wang et al.~\cite{Wang1997} added fluid flow by application of two-phase flow theory but considered only small polymer volume fractions and small gradients in the volume fraction. Durning and Morman \cite{Durning1993} also used continuity equations to describe the flow of solvent and solution in the gel, but used a diffusion approximation with a constant diffusion coefficient to determine the fluid motion. More recently, Wolgemuth et al.~\cite{Wolgemuth2004} extended these theories to study the swelling of a polyelectrolyte gel. However, these current state of the art models do not couple the binding/unbinding of the ions to the network micro-structure, an idea which is critical in describing the bulk mechanical properties of the polymer \cite{Mori2013, English1996}.

The purpose of this paper is to provide a new comprehensive model detailing the swelling kinetics of mucin-like ionic gels and use it to calibrate the rheological data of spherically symmetric mucin granules, exocytosed from the goblet cells of CF-patients versus those released from the non-CF individuals. The novel feature of this model is the coupling of the dynamical motion of the swelling gel with the ionic binding to the polymer network. We use this model to show how the altered kinetics of the CF-infected mucus depends on the electro-chemical environment of the swelling media as well as the rheological properties of the polymer network. The next section presents the details of this model (\S \ref{sec:model}), including equations of motion (\S \ref{subsec:EoM}) and chemistry of binding reactions (\S \ref{subsec:IonChem}). Sections \S \ref{subsec:LinAna} and \S \ref{subsec:ParaEst} outline the linearized analysis of spherically symmetric swelling gels and methods to estimate model parameters, respectively. The results pertaining to the swelling kinetics and the equilibrium configuration of these ionic gels under different chemical stimuli are presented in \S \ref{sec:results}. We conclude with a brief discussion of the implication of these results on the pathophysiology of the infected mucus.

\section{Multi-species, multi-phase mucus-gel model} \label{sec:model}
The polymer gel is modeled as a multi-component material, consisting of $k$ different types of particles.  Specifically, this material consists of solvent particles, polymers, and several small molecular ion species. The polymer is assumed to be made up of monomers (i.e., charge units), denoted as~${\rm M}$, each of which carries a single negative charge. The positively charged ions in the solvent are Hydrogen (H$^+$), Sodium (Na$^+$) and Calcium (Ca$^{2+}$). The negatively charged ions could include Hydronium (OH$^-$) and Chloride (Cl$^-$). Because the negatively charged ions are assumed to be not involved in any binding reactions with the gel, acting only as counterions to positive charges, we identify these ions by the name chloride. The binding reactions of the positively charged ions with the monomers are
 \ben
{\rm M^-}+{\rm H^+}\xrightleftharpoons[\text{k$_{-h}$}]{\text{k$_h$}} {\rm HM} ,~{\rm M^-}+{\rm Na^+} \xrightleftharpoons[\text{k$_{-n}$}]{\text{k$_n$}} {\rm NaM},~{\rm M^-} + {\rm Ca^{2+}}\xrightleftharpoons[\text{k$_{-c}$}]{\text{k$_c$}} {\rm MCa^+},~{\rm M^-} +{\rm MCa^+} \xrightleftharpoons[\text{k$_{-x}$}]{\text{k$_x$}} {\rm M_2Ca}, \label{eqn:chem}
\een
where k$_C$ and k$_{-C}$, for $C = h, n, c, x$, are the binding and the unbinding rates, respectively. We assume that all the binding sites/charge sites are identical although the binding affinities for the different ions are different. The species ${\rm M_2Ca}$ are cross-linked monomer pairs, and the species ${\rm M}^-$, ${\rm MCa}^+$, ${\rm NaM}$ and ${\rm HM}$ are different monomer species, all of which move with the same polymer velocity. The ion species are freely diffusible, but because they are ions, their movement is restricted by the requirement to maintain electroneutrality. Finally, because a small amount of water dissociates into hydrogen and hydronium, we are guaranteed that there are always some positive and negative ions in the solvent.

\subsection{Equations of motion and interface conditions} \label{subsec:EoM}
Suppose we have some volume $V$ of a mixture comprised of $k$ types of particles (including polymer, solvent and ion species) each with particle density (number of particles per unit volume), $n_j({\bf x}, t)$ and particle volume, $\nu_j$, moving with velocity, ${\bf v}_j({\bf x}, t)$, $j = 1,\ldots, k$. The subscripts ($1, 2$) denote the polymer and the solvent phase, respectively. We denote these phases with subcripts ($p, s$) respectively. The volume fractions for polymer and solvent are $\theta_p = \nu_p n_p$ and $\theta_s=\nu_s n_s$. Conservation of polymer implies
%
%
%
\ben
\frac{\partial \theta_p}{\partial t} + \nabla\cdot({\bf v}_p\theta_p)=0.\label{eq:eq_motion1}
\een
We assume that the other molecular species do not contribute significantly to the volume.  Therefore, the volume fractions, $\theta_p+\theta_s=1$. It follows from (\ref{eq:eq_motion1}) and a similar conservation argument for solvent that 
\ben
\nabla\cdot (\theta_s{\bf v}_s+\theta_p{\bf v}_p) = 0. \label{eq:1a}
\een
The motion of the polymer and solvent phase of this multi-component mixture is governed by the Stokes equation for Newtonian fluid, which are
 \ben
\nabla\cdot( \theta_p \sigma_p({\bf v}_p)) - \xi \frac{\theta_s}{\nu_s} \phi_p ({\bf v}_p-{\bf v}_s) - \frac{\theta_p}{\nu_p} \nabla  \mu_p =0, \label{eq:motionVp}
 \een
and
\ben
 \nabla\cdot(  \theta_s \sigma_s({\bf v}_s)) -\xi\frac{{\theta_s}}{\nu_s}\phi_p  ({\bf v}_s-{\bf v}_p) - \frac{\theta_s}{\nu_s}\sum_{j\ge 3}  \hat{\phi}_j \nabla \mu_j -   \frac{\theta_s}{\nu_s}\nabla  {\mu_s} =0, \label{eq:motionVs}
\een
where $\sigma_j({\bf v}) = \frac{\eta_j}{2} (\nabla  {\bf v} + \nabla  {\bf v}^T) + \lambda_j I\nabla\cdot {\bf v}$, are the viscous stresses, $\eta_j>0$ and $\lambda_j$ are the  viscosities, $\xi$ is the drag coefficients. $\phi_p = \nicefrac{n_p}{n_p+n_s}$, $\phi_s = \nicefrac{n_s}{n_p+n_s}$, $\phi_j = \nicefrac{n_j}{\sum_{i\ne 1,2}n_i}$ for $j\ge 3$ (assuming that the ions are dissolved in the solvent), are the polymer, solvent and ion species per total solvent particle fractions, respectively. The first and second terms in \textsc{pde} (\ref{eq:motionVp}, \ref{eq:motionVs}) represents the viscous forces and drag forces due to the friction between the two phases, respectively. The third term in \textsc{pde} (\ref{eq:motionVs}) represents the force due to the dissolved counter-ions in the solvent (osmotic effect), while the last terms in \textsc{pde} (\ref{eq:motionVp}, \ref{eq:motionVs}) are the chemical forces due to the respective phases. The chemical potentials are given by
\begin{align}
\mu_p &= {\it k}_B T M_p + z_m \Phi_e + \nu_p P, \nonumber \\
%
\mu_s &= {\it k}_B T M_s + \nu_s P,
\end{align}
where $M_p$ and $M_s$ represent the entropic contribution to the respective chemical potentials (described later in this section), ${\it k}_B$~is the Boltzmann constant, $T$~is temperature, $P$~is pressure, $z_m$ is the average charge per monomer (defined later in \S \ref{subsec:IonChem}), $\Phi_e$ is the electric potential arising from the unbalanced charges. For the ion species, the particle volume, $\nu_j$, is effectively zero so that
\ben
\mu_j = k_BT\left( \ln\phi_j+1-2\sigma_I  \right) + z_j\Phi_e, \qquad j\ge 3,\label{eq:ion_Chem_Pot}
\een
where $z_j$ is the charge on the $j$th ionic species. In calculating particle fractions, $\phi_j$, we assume that the particle density of the ions is insignificantly small compared to the particle density of polymer and solvent, $\sum_{j\ge3} n_j \ll n_s+n_p$. The ion species satisfy the force balance
\ben 
\xi_j  n_j ({\bf v}_s-{\bf v}_j)-n_j\nabla \mu_j=0, \qquad j\ge 3. \label{eq:ion_mom}
\een
Finally, since there is a free moving-edge to the gel, on one side of which (inside the gel) $\theta^-_p = \theta_p$, and on the other side of which (outside the gel)  $\theta_p^+ = 0$, $\theta_s^+ = 1$, implying that there is no polymer outside the gel. The superscripts ($+, -$) denote regions inside and outside the gel, respectively. The interface conditions are 
\ben
  \sigma_p({\bf v}_p^-)  {\bf n}=   \frac{k_BT} {\nu_m}\left( M_p^-+   z_m\Psi_e + \nu_p\frac{P}{k_BT} \right){\bf n}  ,\label{eq:edge_Conditions_5}
 \een
 and
\ben
 \left(\sigma_s({\bf v}_s^+) - \sigma_s({\bf v}_s^-)\right){\bf n}   = \frac{k_BT} {\nu_s}\left(M_s^+ - M_s^--\sigma_I^+ +\sigma_I^- -\nu_s\frac{P}{k_BT}\right) {\bf n} ,\label{eq:edge_Conditions_6}
 \een
where 
\begin{align}
M_p &= \frac{1}{N} \ln \phi_p + \left(\frac{1}{N}-1\right)\phi_s + \frac{T_0}{T} \left( \frac{\chi}{2}\phi^2_s + \mu^p_0 \right), \nonumber \\
M_s &= \ln \phi_s + \left(1-\frac{1}{N}\right)\phi_p - \sigma_I + \frac{T_0}{T} \left( \frac{\chi}{2}\phi^2_p + \mu^s_0 \right). \label{eq:Ms}
\end{align}
$N$ is the number of monomers per polymer-chain and $T_0$ is a reference temperature. The hydrostatic pressure is $P^- = P$, $P^+=0$ and the normalized electrostatic potential ($\Psi_e = \nicefrac{\Phi_e}{k_B T}$) is, $\Psi_e^- = \Psi_e$, $\Psi_e^+ = 0$. The normal to the free surface is denoted by {\bf n}. The term $\sigma_I$ in the solvent chemical potential (Eq. (\ref{eq:Ms})) is the total ion particle fraction ($\sigma_I = \sum_{j \ge 3} \nicefrac{n_j}{n_s}$) and represents osmotic pressure as characterized by van't Hoff's law. The quantities $\chi, \mu^p_0, \mu^s_0$ are the Flory interaction parameter, standard free energies for pure polymer and pure solvent respectively. These are related to the cross-link fraction, $\alpha$ (the fraction of monomers bound with calcium),
\begin{align}
\chi &= z(\epsilon_1 + \epsilon_2) - 2\left(1-\frac{1}{N}\right)\epsilon_1 - \epsilon_1 \alpha, \nonumber \\
\mu^p_0 &= -\epsilon_1\frac{ z }{2} +\epsilon_4\left(1-\frac{1}{N}\right) + \frac{1}{2}\epsilon_3\alpha, \nonumber \\
\mu^s_0 &= -\epsilon_2\frac{z}{2}. \label{eq:Flory_coeffs}
\end{align}
$\epsilon_i~(i=1, \ldots, 4)$ are the nearest neighbor interaction energies parameters of the various monomer-monomer and monomer-solvent pairs and $z$ is the number of interaction sites on a lattice (i.e., coordination number) \cite{Keener2013}. Eliminating $P$ from Eqns. (\ref{eq:edge_Conditions_5}, \ref{eq:edge_Conditions_6}) we find a single interface condition
 \ben
 \left(\sigma_p({\bf v}_p^-)-\sigma_s({\bf v}_s^-)+  \sigma_s({\bf v}_s^+)\right){\bf n}= \Sigma_{\text{net}}{\bf n}, \label{eq:ic}
 \een
 where $\Sigma_{\text{net}}$ is the net swelling pressure at the interface,
  \ben
 \frac{\Sigma_{\text{net}}}{k_BT}=  \frac{M_p^-}{\nu_m}-\frac{M_s^-}{\nu_s}+ \frac{T_0}{T}\frac{\mu_s^0}{\nu_s}+  \frac{z_m}{\nu_m}\Psi_e +\frac{\sigma_I^-}{\nu_s} -\frac{\sigma_I^+}{\nu_s } .        \label{eq:net_swelling}
 \een
In the above equation, the fourth term represents swelling pressure coming from any electric charge on the monomers, referred as the Donnan pressure ($z_p \Psi_e$ is the corresponding Donnan potential) whereas the last two terms represents the osmotic swelling pressure coming from the difference between the concentrations of ions dissolved in the gel and those dissolved in the bath. Sircar et al. \cite{Keener2013} derived these equations of motion and the interface conditions using the standard variational arguments to minimize the rate of work dissipated within the polymer and the solvent. Unlike previous theories, this model accurately captures how the binding/unbinding of the dissolved ions influences the motion of the swelling gel. The chemistry of the dissolved ions and the charged polymer species is described next.

%
%
\subsection{Ionization chemistry} \label{subsec:IonChem}
This section formulates a mathematical model to represent how the chemical species move and react. Let the concentrations per total volume of the polymer species be denoted by $x = [{\rm M_2Ca}]$, $m = [{\rm M^-}]$, $v=[{\rm NaM}]$,  $w=[{\rm MCa^+}]$ and $y=[{\rm HM}]$, with the total monomer concentration
 \ben
 m_T = m + 2x + w+v+y.\label{eq:47}
 \een
The concentrations per solvent volume of the ion species are denoted as $c=[{\rm Ca^{2+}}]$, $n = [{\rm Na^+}]$, $h = [{\rm H^+}]$,  and $c_l = [{\rm Cl^-}]$. With concentrations expressed in units of moles per liter, the relationship between ion particle fractions $\phi_j$ and concentrations $c_j$ is $\phi_j = \nu_sN_A c_j$, where $N_A$ is Avagadro's number. To describe the chemical reactions, we use the law of mass action. Since all the monomer species are advected with the polymer velocity ${\bf v}_p$, the monomer species evolve according to
\ben
\frac{\partial j}{\partial t} + \nabla \cdot ({\bf v}_p j) = R^+_j - R^-_j, \qquad j=x,v,w,y \label{eq:Nerst-Planck}
\een
where $R^+_j, R^-_j$ are the forward and backward rates for the monomer binding reactions in Eqn.~(\ref{eqn:chem}), respectively. The monomer concentration, $m$, is obtained from Eqn.~(\ref{eq:47}) and $m_T = \nicefrac{\theta_p}{\nu_p N_A}$. Under the assumption of fast chemistry, we set the right hand side of the \textsc{pde} (\ref{eq:Nerst-Planck}) to zero, which reduces into the following set of equations for each of the monomer species
\ben
  w=\frac{\theta_s}{ K_c\phi_s^2}  mc, \quad v= \frac{\theta_s}{K_n\phi_s^2}  mn, \quad y=  \frac{\theta_s}{K_h\phi_s^2} mh, \quad x= \frac{\theta_s}{4K_c^2\phi_s^4}  m^2c,\label{eq:chem_bal}
 \een
where $K_c = \nicefrac{ k_{-c}}{ {k}_c}$, $K_n = \nicefrac{ k_{-n}}{{ k}_n}$, $K_h = \nicefrac{ k_{-h}}{ {k}_h}$, $K_x = \nicefrac{k_{-x}}{ {k}_x} = 4K_c$. The above expressions (\ref{eq:chem_bal}), assume that the unbinding (dissociation) reactions are ionization reactions that require two ``units" of solvent. We take the unbinding reaction rates to be $ k_{-C}\phi_s^2$, for $C = c, x, n, h$ and because calcium is a divalent ion, $2k_x=k_{c}$ and $k_{-x} = 2k_{-c}$.

Similarly, under the assumption of fast diffusion and chemistry, the law of mass action for the ion species reduces to
\ben C = C_b e^{-z_C \Psi_e} 
\label{eq:ion_bal3c} \een
with symbol $C = c, h, n, c_l$,~$z_n= z_h=1, z_c=2$ and $z_{c_l} = -1$. The subscript `b' denotes the corresponding bath concentrations \cite{Keener2013}. The electrostatic potential, $\Psi_e$, is determined by the electroneutrality constraint inside the gel, namely,
 \ben  (2c + n + h-c_l)\theta_s + z_m m_T=0, \label{eq:el_bal} \een
where $z_m$ is the average residual charge of the unbound monomers which depends on the amount of binding with ions,
\ben z_m m_T = w - m. \een
Since both the electrostatic potential and polymer particle fraction are assumed to be zero outside the gel, electroneutrality in the bath requires that
%
%
\ben
2c_b + n_b + h_b - {c_l}_b = 0. \label{eq:bath}
\een
Finally, the crosslink fraction (or the fraction of monomers bound with calcium) is $\alpha = \frac{x}{m_T}$, where Eqns.~(\ref{eq:47}, \ref{eq:chem_bal}) define the concentrations, $m_T, x$, respectively.

In summary, the dynamical motion of a freely swelling mucus-gel is modeled by the system of equations including the mass conservation \textsc{pde}~(\ref{eq:eq_motion1}), total volume conservation \textsc{pde}~(\ref{eq:1a}) together with force balance \textsc{pdes}~(\ref{eq:motionVp}-\ref{eq:motionVs}) and interface condition Eqn.~(\ref{eq:ic}), subject to the constraints Eqn.~(\ref{eq:47}) (monomer conservation), Eqn.~(\ref{eq:ion_bal3c}) (ion motion) and Eqn.~(\ref{eq:el_bal}) (electroneutrality).
%
%
\subsection{Linearized analysis of spherically symmetric swelling gels} \label{subsec:LinAna}
We now consider the swelling kinetics of a radially symmetric mucus gel using a linearized analysis of the governing equations, described in the previous two sections. The volume fraction $\theta_p$ is non-zero on the domain $0 \le r < R(t)$, with $R$ (maximum radius of the sphere) a function of time. Since only the finite solutions of the governing equations are of interest (in particular, finite solutions of the force balance, \textsc{pdes}~(\ref{eq:motionVp}, \ref{eq:motionVs})) inside this domain, we assume that $v_p = v_s = 0$ at $r = 0$. The volume conservation, \textsc{pde}~(\ref{eq:1a}), implies the constraint $\theta_p v_p + \theta_s v_s = 0$ throughout the domain. Using this constraint, we reduce the two force balance equations into a single equation (by multiplying \textsc{pde}~(\ref{eq:motionVp}) by volume fraction $\theta_s$ and \textsc{pde}~(\ref{eq:motionVs}) by volume fraction $\theta_p$ and subtracting)
\begin{align}
&\theta_s \nabla \cdot \left[\theta_p \left(\eta_p\frac{\partial {\bf v}_p}{\partial r} + \frac{\lambda_p}{r^2} \frac{\partial }{\partial r}r^2 {\bf v}_p\right) \right] + \theta_p \nabla \cdot \left\{\theta_s \left[ \eta_s \frac{\partial }{\partial r}\left(\frac{\theta_p {\bf v}_p}{\theta_s}\right) + \frac{\lambda_s}{r^2} \frac{\partial }{\partial r}\left(\frac{r^2 \theta_p {\bf v}_p}{\theta_s}\right)\right] \right\} 
\nonumber \\
&+ \xi {\bf v}_p \theta_p \left(\frac{\phi_p}{\nu_s} + \frac{\phi_s}{\nu_p}\right) -\theta_p \theta_s \nabla \left(\Sigma_{\text{net}}\right) = 0, \label{eq:mod-force}
\end{align}
whereas the interface condition (Eqn.~(\ref{eq:ic})) reduces to
\ben
\eta_p \frac{\partial {\bf v}_p}{\partial r} + \eta_s \frac{\partial }{\partial r}\left( \frac{\theta_p {\bf v}_p}{\theta_s}\right) + \frac{\lambda_p}{r^2}\frac{\partial }{\partial r}r^2 {\bf v}_p + \frac{\lambda_s}{r^2}\frac{\partial }{\partial r} \left(\frac{r^2 \theta_p {\bf v}_p}{\theta_s}\right) = \Sigma_{\text{net}}  \qquad \textrm{at}~~r=R(t). \label{eq:mod-ic}
\een
Note that $\nabla(\frac{\mu_p}{\nu_m} - \frac{\mu_s}{\nu_s}) = \nabla (\Sigma_{\text{net}})$, where $\Sigma_{\text{net}}$ is the interface swelling pressure defined in Eqn. (\ref{eq:net_swelling}). The velocities inside the gel are ${\bf v}^-_{p/s} = {\bf v}_{p/s}$ and outside the gel are ${\bf v}^+_{p/s}=0$. Next, we assume that the linearized velocity $v_p$ is small (the equilibrium velocity $v^*_p=0$) and that $\theta_p=\theta^*_p + \delta\theta_p$, where $\theta^*_p$ is the equilibrium polymer volume fraction. Sircar et al. \cite{Keener2013} gives details on the  equilibrium solution. Since this is a moving boundary problem, it is convenient to map the domain $0 \le r < R(t)$ onto the fixed domain $0 \le y < 1$ by making the change of variables $r = R(\tau)y$ and $t=\tau$. Further, we seek space-time variable separated solutions of the form $v_p = a(\tau) f_1(y)$ and $\delta \theta_p = a(\tau) f_2(y)$. Under these assumptions the reduced force balance, \textsc{pde}~(\ref{eq:mod-force}) is
\ben
f''_1 + \frac{2(\eta_e+2\lambda_e)R}{(\eta_e+\lambda_e)y} f'_1 + \left[\frac{2\lambda_e}{(\eta_e+\lambda_e)y^2} + \frac{\xi R}{\eta_e+\lambda_e}\left(\frac{\phi^*_s}{\nu_p} + \frac{\phi^*_p}{\nu_s}\right) \right]f_1 = -\theta^*_s R \Sigma_{\text{net}}^{\theta}(\theta^*_p) f'_2, \label{eq:linear_fb}
\een
and the interface condition, Eqn.~(\ref{eq:mod-ic}), is
\ben
(\eta_e + \lambda_e)f'_1 + \frac{2\lambda_e}{y} f_1 = \theta^*_s \Sigma_{\text{net}}^{\theta}(\theta^*_p) f_2 \qquad \textrm{at}~~y=1.
\een
The supercripts $( ' )$ and $(^{\theta})$ denote the derivative of the functions with respect to the variables $y$ and $\theta$, respectively. The net shear and bulk viscosities are given by $\eta_e = \theta^*_s\eta_p + \theta^*_p \eta_s$ and $\lambda_e = \theta^*_s\lambda_p + \theta^*_p \lambda_s$. $\phi^*_p, \phi^*_s$ are the equilibrium particle fractions for the polymer and solvent, respectively. For a finite solution of Eqn.~(\ref{eq:linear_fb}), it is assumed that $f_1(0)=0$. Further, we assume that $f'_2 = f_1$. These assumptions reduce Eqn.~(\ref{eq:linear_fb}) into the homogeneous, spherical Bessel differential equation whose solutions have the form $f_1 = y^{\gamma} J_n(\beta y)$, where $\gamma=\frac{1}{2} - \frac{(\eta_e+2\lambda_e)R}{\eta_e+\lambda_e}$, $\beta=\sqrt{\frac{\xi R}{\eta_e+\lambda_e}(\frac{\phi^*_s}{\nu_p} + \frac{\phi^*_p}{\nu_s})+\theta^*_s R \Sigma_{\text{net}}^{\theta}(\theta^*_p)}$ and $n = \sqrt{|\gamma^2 - \frac{2\lambda_e}{\eta_e+\lambda_e}|}$. The Bessel functions, $J_n$, are of the first kind of order $n$. Since there are several solutions satisfying the condition $f_1(0)=0$, we choose the solution with the lowest order $n=n_{\text{min}}$, which determines the equilibrium radius of the swelling gel, $R_f = R(\tau \rightarrow \infty)$, 
\ben
R_f = \left( \frac{1}{2} - \sqrt{n^2_{\text{min}}+\frac{2\lambda_e}{\eta_e+\lambda_e}} \right) \left(\frac{\eta_e+\lambda_e}{\eta_e+2\lambda_e} \right). \label{eq:finalRad}
\een
The variable-separable, linearized form of solution (linearized near equilibrium) for mass conservation, Eqn.~(\ref{eq:eq_motion1}), gives
\ben
\frac{1}{a} \frac{\partial a}{\partial \tau} = -\frac{1}{f_2} \frac{\theta_p^*}{R_f y^2} (y^2 f_1)' = -\frac{1}{\tau_{ch}}, \label{eq:linear_mass_conserv}
\een
where $\tau_{ch}$ is a constant (to be determined later). The first and the last part of Eqn.~(\ref{eq:linear_mass_conserv}) implies that $a(\tau) = e^{-\tau / \tau_{ch}}$. Because the polymer is conserved (neither created nor destroyed) the velocity of the moving boundary must be the same as the gel velocity at the boundary, $v_p(y=1)$, 
\ben
\frac{\partial R}{\partial \tau} = v_p(y=1) = e^{-\tau / \tau_{ch}} f_1(1). \label{eq:Rtau}
\een
The solution to the above equation is numerically computed via Matlab ODE solver ode15, with trivial initial conditions. In particular, we note that if $R \approx R_f$, then $f_1(1) \approx$ constant and in this case Eqn. (\ref{eq:Rtau}) can be solved exactly, i.e., $R(\tau) = R_f (1-e^{-\tau / \tau_{ch}})$. This expression is identical to the radial expansion of the linearized form given by Tanaka {\it et. al} \cite{Tanaka1979} in their kinetic theory of swelling hydrogels. Finally, the time constant of gel-swelling towards the equilibrium size (Eqn. (\ref{eq:linear_mass_conserv})), is
\ben
\tau_{ch} = \frac{f_2 R_f y^2}{\theta^*_p (y^2 f_1)'} \Big|_{y=1}, \label{eq:charTime}
\een
which defines the diffusivity for spherical gels, $D = \nicefrac{R^2_f}{\tau_{ch}}$.
%
%
\subsection{Parameter Estimation} \label{subsec:ParaEst}
\begin{figure}
\centering
\subfigure[{$pH=7.2$, [Ca$^{2+}$]$_b=1$\,mM, [Na$^+$]$_b=0$}]{\includegraphics[scale=0.5]{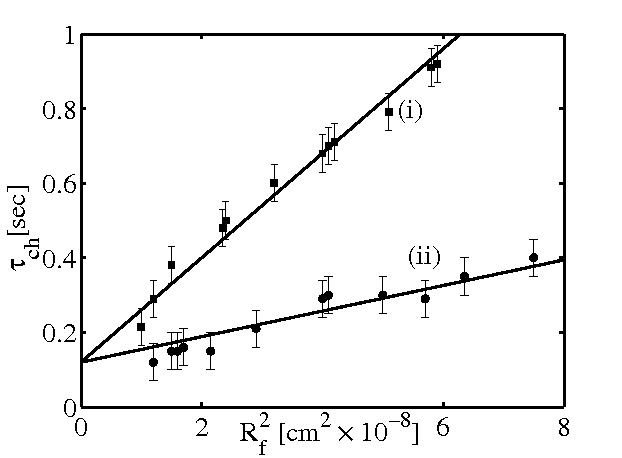}}
\subfigure[{$pH=7.0$, [Ca$^{2+}$]$_b=4$\,mM, [Na$^+$]$_b=140$\,mM}]{\includegraphics[scale=0.5]{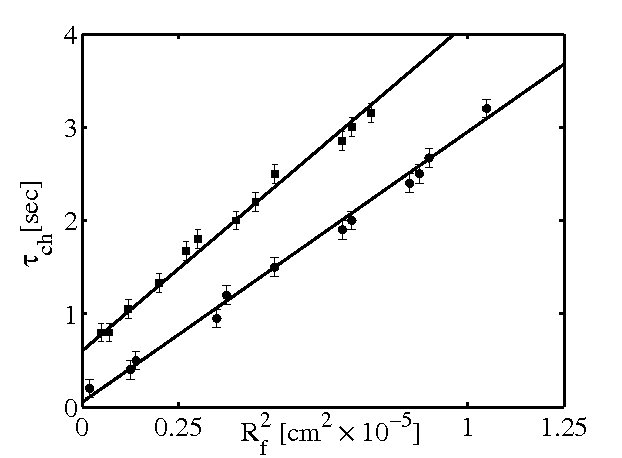}}
\caption{Experimental data of the final size squared ($R^2_f$) versus characteristic swelling time~($\tau_{ch}$) for the two {\it in vitro} electro-chemical conditions mentioned in \S 2.4. For WT mucus the data points are represented by ($\CIRCLE$) and for CF-infected mucus these points are shown by ($\blacksquare$). The temperature of the experiments was set at $37^\circ$C. The model output (i.e., solid lines) closely fits the experimental data points. Numerical simulations are done for the data represented by the points {\bf (i)} and {\bf (ii)}. These are shown in Fig. \ref{fig:Fig3}.}\label{fig:Fig2}
\end{figure}

Two sets of data are used to calibrate the model for the kinetics of spherically symmetric swelling mucus gels. The first experiment, conducted by Verdugo et al., measures the diffusivity data ($R^2_f$ versus $\tau_{ch}$ values) of the swelling kinetics of exocytotic mucin granules from four (human) CF-patients and three healthy individuals at $pH=7.2$, $T=37^\circ$C and [Ca$^{2+}$]$_b=1$mM as well as [Ca$^{2+}$]$_b=2.5$mM \cite{Verdugo1998}. In the second experiment, performed by Kuver et al., kinetic data was collected from cultured gallbladder epithelial cells from wild-type and CF-infected mice at $pH=7.0$, $T=37^\circ$C, [Ca$^{2+}$]$_b=4$mM and [Na$^+$]$_b=140$mM \cite{Kuver2006}. The microscopic composition of the mucus in both of these experiments were reported identical. Fig.~\ref{fig:Fig2}a and Fig.~\ref{fig:Fig2}b presents the results from these two experiments, respectively.

Table~\ref{tab:Table1} lists the values of the parameters used in our numerical calculations. The constants in the model are the monomer particle volume, $\nu_p$, the solvent particle volume, $\nu_s$, the coordination number of the polymer lattice, $z$, and the nearest neighbor interaction energies,~$\epsilon_i$ (Eqn.~(\ref{eq:Flory_coeffs})), shear and bulk viscosity coefficients of solvent (i.e., water), $\eta_s, \lambda_s$, respectively, at reference temperature. The undetermined constants are the binding affinities of the various cations with the gel, $K_h, K_c, K_n$ (introduced in Eqn.~(\ref{eq:chem_bal})) and the polymer viscosity coefficients, $\mu_p, \lambda_p$, and the drag coefficient, $\xi$ (Eqn.~(\ref{eq:motionVp})).
\begin{table}[htbp]
\centering
\caption{Constant parameters common to all the numerical simulations. The reference temperature for the viscosity coefficients and solubility parameters is fixed at $T_0=25^\circ$C.}\label{tab:Table1}
\begin{tabular}{|l|l|l|l|l|}
\hline
 Constants & Value & Units & Source\\
\hline
Repeat unit per chain ($N$) & 266 & -- & \cite{Villar2007} \\
Molecular volume of mucus ($\nu_p$) & 5 $\times$ 10$^{-20}$ & m$^3$ & \cite{Villar2007} \\
Molecular volume of water ($\nu_s$) & 2 $\times$ 10$^{-23}$ & m$^3$ & \cite{Villar2007} \\
Shear viscosity of water ($\eta_s$) & 8.88 $\times$ 10$^{-4}$ & Pa s & \cite{Holmes2011} \\
Bulk viscosity of water ($\lambda_s$) & 2.47 $\times$ 10$^{-3}$ & Pa s & \cite{Holmes2011} \\
Hildebrand solubility ($\delta_p$) & 1.0928 ($\alpha=0$), 1.3258 ($\alpha=1$) & MPa$^{1/2}$ & \cite{Mimura1989} \\
Hildebrand solubility for water ($\delta_s$) & 48.07 & MPa$^{1/2}$ & \cite{Barton1990} \\
\hline
\end{tabular}
\end{table}

The radially symmetrically swelling mucus gel has a 3-D configuration, which suggests that we choose the coordination number, $z=6$, mimicking the 3-D structure of the polymer lattice~\cite{Flory1953}. The standard free energies, ${k}_B T_0 \mu^0_p$ and ${k}_B T_0 \mu^0_s$ and the energy interaction parameters, $\epsilon_i$~(Eqn.~(\ref{eq:Flory_coeffs})) are found from the Hildebrand solubility data, $\delta_i$ \cite{Barton1990}. The values for the solubility data for materials mimicking mucus glycoproteins are given in a study by Mimura \cite{Mimura1989}. The fully un-crosslinked (no Ca$^{2+}$ binding) and fully crosslinked (Ca$^{2+}$ bound) states are denoted by $\alpha=0$ and $\alpha=1$, respectively. The standard free energy is the energy of all the interactions between the molecule and its neighbors in a pure state that have to be disrupted to remove the molecule from the pure state. The relation between the standard free energies and the solubility parameters (Table~\ref{tab:Table1}) are
\begin{align}
-{\it k}_B T_0 \mu^0_s &= \nu_s\delta^2_s \nonumber \\
-{\it k}_B T_0 \mu^0_p &= \nu_m\delta^2_p, \label{eq:HSP}
\end{align}
where $\nu_s=2 \times 10^{-23}$ cm$^3$ is the volume of one molecule of water at reference temperature, $T_0 = 298$K. The negative sign in Eqn.~(\ref{eq:HSP}) indicates that ${\it k}_B T_0 \mu^0_p, {\it k}_B T_0 \mu^0_s < 0$, since they are the interaction energies. Using the relations in Eqn.~(\ref{eq:HSP}), these values are fixed at $\epsilon_1=4.84,~\epsilon_2=3.74,~\epsilon_3=-13.70,~\epsilon_4=0$. The reference temperature of the experiments was fixed at $T_0=25^\circ$C.

The volume of a mucin oligomer/multimer chain is calculated by mathematically modeling a chain comprising $N$ freely jointed cylindrical segments of length $l$ and width $d$, where $N=L/l$ is the number of Kuhn segments (or effective rigid cylindrical segments which determines the different conformations a chain can have), and $L$ is the end-to-end length of the chain. For a gel forming mucin (e.g., human \textsc{muc5ac}, used in the experimental data to calibrate our model), $l \approx 0.03\,\mu$m, $L \approx 8\,\mu$m and $N=8/0.03 \approx 266$ \cite{Bansil2006, Verdugo1990}. The radius of gyration, $R_g$, (defined as the average distance from all Kuhn segments to the center of mass of the chain) and the pervade volume, $V$, (i.e., the approximate spherical volume of a sphere with radius $R_g$) is \cite{Doi1986}
\ben
R_g \approx \frac{l^{0.8} \times d^{0.2} \times N^{0.588}}{2.45}, \qquad V \approx 4 \times \frac{\pi}{3} \times R^3_g.
\een
Substituting values, the volume $V (=\nu_p) \approx 0.05\,\mu$m$^3$. 

The undetermined parameters, namely the binding affinities $K_h, K_n, K_c$ and the rheological coefficients $\eta_p, \lambda_p, \xi$  are computed by minimizing a nonlinear least-square difference function between the experimental values of diffusivity data ($R^2_f$ versus $\tau_{ch}$) and the corresponding model output (Eqns. (\ref{eq:finalRad}, \ref{eq:charTime})), implemented via the \textsc{matlab} non-linear least-square minimization function lsqnonlin. These values are found as $\log_{10}(K_n)=-2.27$,\,$\log_{10}(K_h)=-3.65$,\,$\log_{10}(K_c)=-3.12$,\,$\eta_p=0.11$,\,$\lambda_p=0.31$,\,$\xi=1.97$\,(wild-type mucus) and $\log_{10}(K_n)=-2.55$,\,$\log_{10}(K_h)=-3.98$,\,$\log_{10}(K_c)=-7.12$,\,$\eta_p=1.02$,\,$\lambda_p=2.87$,\,$\xi=0.21$\,(CF-infected mucus). The closeness of fit between the model (highlighted by the solid lines) and the experimental data points is shown in Fig~\ref{fig:Fig2}. The error bars represent the maximum and minimum deviation from the sample points and set at 5\% margin of error. Notice the linear relationship between $R^2_f$ and $\tau_{ch}$, predicted by the experiments and accurately captured by the model.

\section{Results and discussion} \label{sec:results}
The main idea behind this work is to provide an objective comparison of the swelling properties of normal versus unhealthy mucus, immersed in an extracellular medium with chemically controlled composition. Well established results exist in literature which detail the relationship between the swelling rate, final size of the gel and the swelling time \cite{Tanaka1979,Tanaka1980}. However, these results do not explicitly outline how these and other physiochemical elements influencing the mucus hydration and rheology, depend on the nature of the swelling media. Hence, in \S \ref{subsec:SK}, we explore the relationship between the radial size of the swelling mucus gel and the rheology of the mucus polymer, name the drag and the viscosity coefficients. The effect of the calcium and the sodium ions in the solvent on the equilibrium size of the mucus blob are detailed in \S \ref{subsec:FinalCa} and \S \ref{subsec:FinalNa}, respectively.

\subsection{Swelling kinetics} \label{subsec:SK}
\begin{figure}
\centering
\includegraphics[scale=0.5]{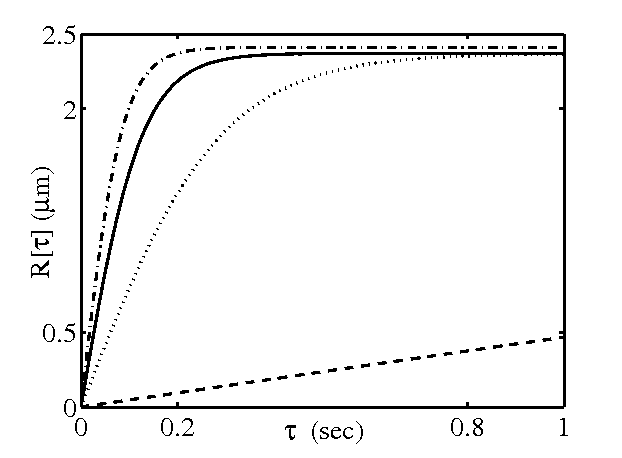}
\caption{Swelling radius, $R(\tau)$, of a spherically symmetric mucin gel versus time, $\tau$, with kinetics governed by Eqn. (\ref{eq:Rtau}), for WT-mucus at calcium concentration C$_b=1.0$\,mM (dash-dot curve), CF-mucus at C$_b=1.0$\,mM (dotted curve), WT-mucus at C$_b=2.5$\,mM (solid curve), CF-mucus at C$_b=2.5$\,mM (dashed curve). The ($R_f, \tau_{ch}$) values for the dash-dot and the dotted curve correspond to the data represented by points {\bf (i)} and {\bf (ii)} in Fig. \ref{fig:Fig2}a, respectively.}\label{fig:Fig3}
\end{figure}

Numerical simulations were performed to outline the differences between the swelling kinetics for WT and CF-infected mucus, by altering the electro-chemical composition of the swelling media. Fig.~\ref{fig:Fig3} presents the radius of the swelling gel, $R(\tau)$, versus time, $\tau$, for WT mucus with calcium bath concentrations, C$_b=1$\,mM and C$_b=2.5$\,mM (the `dash-dot' and `solid' curve, respectively), as well as for CF-infected mucus (highlighted by the `dotted' and `dashed' curve, respectively). The sodium ion concentration and the pH in the bath are fixed at N$_b=0$ and $pH=7.2$. These concentrations correspond to the {\it in vitro} conditions of Verdugo's experiments \cite{Verdugo1998}. 

In particular, note that the expansion of the CF-infected mucus (`dashed' and `dotted' curves, Fig.~\ref{fig:Fig3}) is appreciably slower than WT mucus (`dash-dot' and `solid' curves, Fig.~\ref{fig:Fig3}). This difference in swelling profiles is a consequence of small drag coefficient of the CF-infected mucus gels, relative to the viscosity coefficients (i.e., the ratio $\delta = \frac{\xi}{\theta^*_s(\eta_p+\lambda_p)+\theta^*_p(\eta_s+\lambda_s)}$ is small for CF-infected mucus). Using the rheological parameters, estimated in \S \ref{subsec:ParaEst} (i.e., $\eta_p, \lambda_p, \xi$), the solvent parameters listed in Table \ref{tab:Table1} (i.e., $\eta_s, \lambda_s$) and the equilibrium polymer volume fraction, $\theta^*_p$ (i.e., by solving Eqns. (\ref{eq:ic}, \ref{eq:el_bal}), values shown in Fig. \ref{fig:Fig4}a); we find that $\delta=82.8655, 78.7264$ for WT mucus for concentrations $C_b=1$\,mM and $C_b=2.5$\,mM, respectively while the corresponding values are $\delta=0.0885, 0.0914$ for CF-infected mucus, respectively. Gels with higher viscosities expand at a slower rate and this explains a relatively slow or insufficient hydration due to the defective rheology of the CF-infected mucus. Previous numerical studies by Sircar at al. corroborate these results \cite{Keener2011}. 

The corresponding diffusivity data for the curves in Fig.~\ref{fig:Fig3} are given in Table \ref{tab:Table2}. The diffusivity decreases by 33\% for normal mucus and 88\% for CF-infected mucus, when the Ca$^{2+}$ concentration in the swelling medium is changed from 1 to 2.5 mM, respectively. Further, there is a 76\% reduction in the diffusivity of CF-infected mucus versus the WT mucus in a solvent containing 1\,mM Ca$^{2+}$, in concurrence with Verdugo's swelling experiments \cite{Verdugo1998}. A lower diffusivity of the defective mucus is a consequence of the slow rate of expansion of these gels (or a bigger time constant, $\tau_{ch}$), which again lead to the conclusion that the defective rheology of mucus plays a crucial role in the eventual mucus hydration.
%
\begin{table}[htbp]
\centering
\caption{Diffusivity data for WT and CF-infected mucus at calcium bath concentrations C$_b~=~1$\,mM and C$_b=2.5$\,mM. $R_f, \tau_{ch}$ and $D$ are measured in $\mu$m, sec and cm$^2$/s, respectively. The sodium concentration and the pH in the bath are fixed at N$_b=0$ and $pH=7.2$.}\label{tab:Table2}
\begin{tabular}{|l|l|l|}
\hline
& 1 mM Ca$^{2+}$ & 2.5 mM Ca$^{2+}$\\
\hline
WT & $R_f=2.41$, $\tau_{ch}=0.2$, $D=2.90 \times 10^{-7}$ & $R_f=2.37$, $\tau_{ch}=0.29$, $D=1.93 \times 10^{-7}$ \\
CF & $R_f=2.37$, $\tau_{ch}=0.79$, $D=7.09 \times 10^{-8}$ & $R_f=2.37$, $\tau_{ch}=6.58$, $D=0.85 \times 10^{-8}$ \\
\hline
\end{tabular}
\end{table}

\subsection{Equilibrium configuration versus bath [Ca$^{2+}$]} \label{subsec:FinalCa}
\begin{figure}
\centering
\subfigure[Equilibrium polymer volume fraction, $\theta^*_p$]{\includegraphics[scale=0.5]{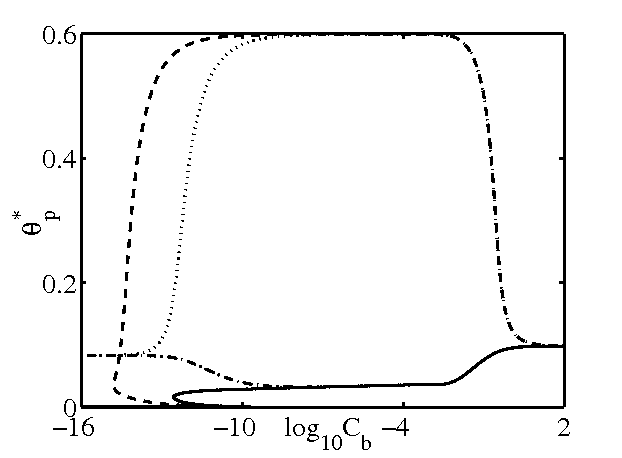}}
\subfigure[Donnan swelling pressure, $z_m \Psi_e$]{\includegraphics[scale=0.5]{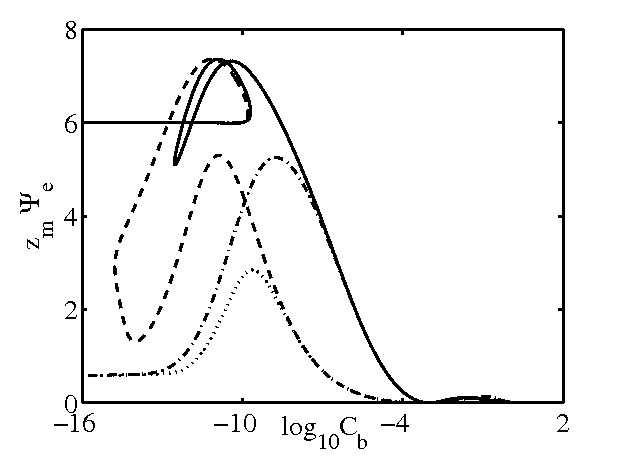}}
\subfigure[crosslink fraction, $\alpha$]{\includegraphics[scale=0.5]{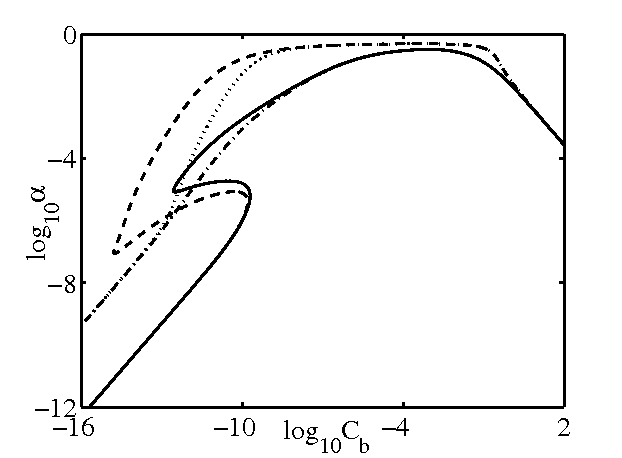}}
\caption{Variables in equilibrium versus calcium bath concentration, $C_b$ (in mol/lt) for WT-mucus at $pH=7.2$ (solid curve), CF-infected mucus at $pH=7.2$ (dashed curve), WT-mucus at $pH=5.0$ (dash-dot curve), CF-infected mucus at $pH=5.0$ (dotted curve). The bath concentration of sodium and the temperature are fixed at $N_b=0$ and $T=37^\circ$C, respectively.}\label{fig:Fig4}
\end{figure}

In the next two sections, we explore the role of the electrolytic composition of the swelling media on the equilibrium configuration of the healthy versus the diseased mucus. Fig.~\ref{fig:Fig4}a,b,c highlight the equilibrium volume fraction,~$\theta^*_p$ (which determines the swelled/deswelled state of the gel), the Donnan potential and the crosslink fraction, at pH$=7.2$ and pH$=5.0$, versus the bath concentration of calcium, respectively. 

The emergence of three different swelling regions are observed in Fig.~\ref{fig:Fig4}. At low bath concentrations of calcium (C$_b<10^{-14}$M), the gels are immersed in a solvent which lacks cations to bind with the charged monomers. This leads to a highly ionized state (i.e., a high average negative charge per monomer) and consequently a higher Donnan potential (Fig. \ref{fig:Fig4}b). Sircar et al. has shown earlier that Donnan potential is the dominant mechanism driving the swelling of ionic gels \cite{Keener2013}. Solvents with lower pH (swelling profiles represented by the `dash-dot' and `dotted' curves, Fig. \ref{fig:Fig4}a) can furnish more H$^+$ ions, leading to a lower Donnan potential which translates to a relatively de-swelled state (i.e., comparing the $\theta^*_p$ values at pH$=5$ versus those at pH$=7.2$).

The hydration properties of the diseased versus healthy mucus are markedly different for small to intermediate calcium concentrations, $10^{-14}$\,M$<$C$_b<10$\,M (Fig.~\ref{fig:Fig4}). In particular, the diseased mucus (swelling profiles shown by the `dashed' and `dotted' curve) shows massive deswelling. The explanation for these swelling features in this region is that there is a complex interplay of ionization via Donnan potential (which favors swelling) and the energy gain due to increased crosslinking (which promotes deswelling) \cite{Keener2013}. Defective mucus gels have higher calcium binding affinity (i.e., $\log_{10}K_c=-7.12$ for CF-infected mucus versus $\log_{10}K_c=-3.12$ for WT mucus, refer \S \ref{subsec:ParaEst} where these binding affinity values are mentioned), This leads to a relatively de-ionized state in the defective gels (or a lower Donnan potential, Fig. \ref{fig:Fig4}b). Consequently, a de-swelled state (mediated by the de-ionization due to calcium crosslinks) dominates.

At sufficiently high calcium ion concentrations (C$_b>10$\,M) the average charge on the monomer is positive rather than negative. The positive charge on the monomer is because in the divalent ion case, the binding of a monomer with an ion converts it from a negatively charged ion, M$^-$, into a positively charged ion, MCa$^+$ (Eqn.~(\ref{eqn:chem})). This has little effect on the overall swelling pressure since at high ion concentrations  the charge on the gel is small relative to the overall number of available ions, and the Donnan potential, vanishes. As a consequence, at sufficiently high $C_b$ the gel (both normal as well as the diseased type) behaves as if it is uncharged. In summary, an increased affinity of the CF-infected mucus to bind with calcium (in physiologically significant concentration range) leads to a highly de-swelled state of the gel. This defective binding property of the diseased mucus partly explains how the altered movement of the electrolytes leads to an insufficient hydration of the gel.

\subsection{Equilibrium configuration versus bath [Na$^+$]} \label{subsec:FinalNa}
\begin{figure}
\centering
\subfigure[Equilibrium polymer volume fraction, $\theta^*_p$]{\includegraphics[scale=0.5]{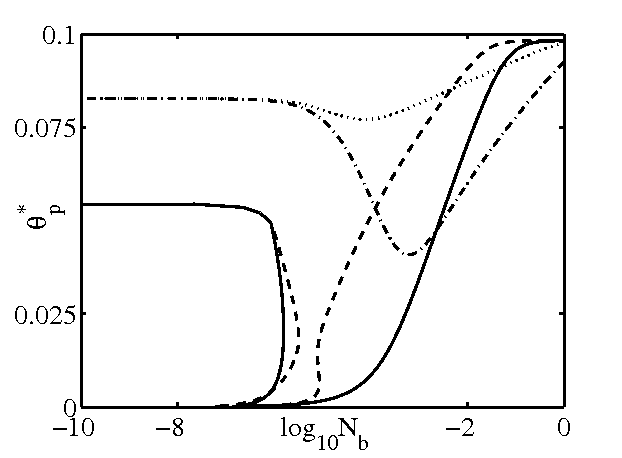}}
\subfigure[Donnan swelling pressure, $z_m \Psi_e$]{\includegraphics[scale=0.5]{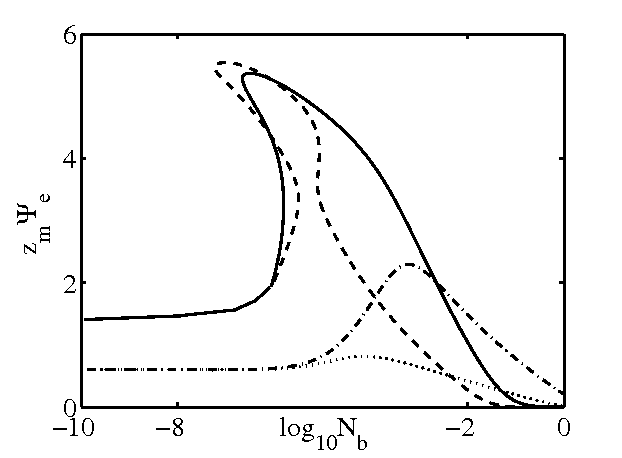}}
\caption{Variables in equilibrium versus sodium bath concentration, $N_b$ (in mol/lt) for WT-mucus at $pH=7.2$ (solid curve), CF-infected mucus at $pH=7.2$ (dashed curve), WT-mucus at $pH=5.0$ (dash-dot curve), CF-infected mucus at $pH=5.0$ (dotted curve). The bath concentration of calcium and the temperature are fixed at $C_b=0$ and $T=37^\circ$C, respectively.}\label{fig:Fig5}
\end{figure}

This section reports the salient features of the equilibrium state of the mucus gel immersed in a bath containing monovalent cations (i.e., Na$^+$, Fig.~\ref{fig:Fig5}). Again, three different swelling transition regimes are seen from these figures: a de-swelled state (i.e., a relatively high polymer volume fraction, $\theta^*_p$) is favored at low (N$_b<10^{-7}$\,M) and high (N$_b>1$\,M) bath concentrations. A higher polymer volume fraction is because of a low Donnan potential in these concentration ranges, which is due to either the furnished H$^+$ ions (at low sodium concentrations) or excessive Na$^+$ ions (otherwise). In small to intermediate sodium concentrations ($10^{-7}$\,M$<$N$_b<10^{-1}$\,M, i.e., approximately in the region N$_b\approx$H$_b$), there is an additional feature with increasing sodium concentration, namely, swelling either gradually ($pH=5.0$ curves, Fig. \ref{fig:Fig5}a) or via phase-transition ($pH=7.2$ curves, Fig. \ref{fig:Fig5}b) followed by de-swelling. The explanation for this swelling feature is a complicated interplay between ionization via hydrogen unbinding (which promotes swelling) and a de-ionization via sodium binding (which promotes de-swelling).

However, the volume transition curves are qualitatively different for the WT and the CF-infected mucus, e.g., notice the hysteretic swelling transition for the WT-mucus (solid curve) versus a double-hysteretic curve for the defective mucus (dashed curve) at $pH=7.2$. These differences, again, arise due to different binding affinities ($\log_{10}K_n=-2.27$,\,$\log_{10}K_h=-3.65$ for CF-infected mucus versus $\log_{10}K_n=-2.55$,\,$\log_{10}K_h=-3.98$ for WT mucus, values listed in \S \ref{subsec:ParaEst}). Thus, the differences in the ion-binding property of the mucus influences the not only the swelling-deswelling volume transitions but also the equilibrium state of the gel.


\section{Conclusions} \label{sec:conclusion}
This paper develops a new, comprehensive, multi-phase, multi-species model to quantify the swelling / deswelling mechanism for mucin gels. This model explains how the final configuration of these gels depends on complex interactions between competing effects that alters the gel ionization, that is the Donnan potential, changes in the bath concentration of ions and their corresponding binding affinity with mucus. Near equilibrium, the radial size of the swelling mucus gel reduces to the well known expression for hydrogel swelling~\cite{Tanaka1979}. The diffusivity, $D$ (which is similar to the one defined by Tanaka's hydrogel theory) accurately characterizes the swelling properties of the mucin network, including the effect of calcium binding on the equilibrium configuration of the gel. The reasons behind the experimental outcomes of deficient mucus hydration for CF-infected patients are examined. In particular, the defective swelling properties of the diseased mucus is the combined effect of an abnormal mucus rheology (i.e., an abnormally low drag relative to the polymer viscosity), an altered electrolytic composition of the extracellular medium (i.e., an increased calcium ion concentration in the airway surface liquid surrounding the mucus) as well as an altered movement of ions in the medium (i.e., a higher affinity of the CF-infected mucus to bind with calcium ion). All these factors favor a condensed state of the gel, resulting in a mucus gel with defective mucociliary transport contributing to the chronic airway infection found in CF patients.

\end{document}